\documentstyle[preprint,prl,aps]{revtex}
\begin{document}
\draft
\preprint{\today}
\title{
 Nonlinear Transport Properties of Quantum Dots }
\author{W.~Pfaff$^{\,a}$, D.~Weinmann$^{\,a,b}$, W.~H\"ausler$^{\,b}$,
B.~Kramer$^{\,b}$, U.~Weiss$^{\,a}$}
\address{ $^{a})$ Universit\"at Stuttgart, II.~Inst.~f.~Theoret.~Physik,
        Pfaffenwaldring 57, 70569 Stuttgart, F.~R.~G. \\
          $^{b})$ Phys.--Techn.~Bundesanstalt, Gruppe~8.1,
        Bundesallee 100, 38116 Braunschweig, F.~R.~G.}
\maketitle
\date{\today}
\begin{abstract}
The influence of excited levels on nonlinear transport properties of a
quantum dot weakly coupled to leads is studied using a master--equation
approach. A charging model for the dot is compared with a quantum mechanical
model for interacting electrons. The current--voltage curve shows Coulomb
blockade and additional finestructure that is related to the excited states
of the correlated electrons. Unequal coupling to the leads causes asymmetric
conductance peaks. Negative differential conductances are predicted due to
the existence of excited states with different spins.
\end{abstract}

Mail/VE V2.0     92196                  read_letter
         1993-06-22         18:13:17     PAGE 2

\pacs{PACS: 72.20.Ht, 73.20.Dx, 73.20.Mf, 73.40.Gk.}
\narrowtext

Periodic oscillations of the conductance through quantum dots that are
weakly coupled to leads \cite{kastner} are well established consequences of
the charging energy of single electrons entering or leaving the dot at
sufficiently low temperatures. They are observed in linear transport as a
function of the carrier density. At bias voltages larger than the
differences between discrete excitation energies within the dot, a
characteristic splitting of the conductance peaks is observed
\cite{johnson,weis92}.
We will demonstrate unambiguously below that this is related to transport
involving the excited states of $n$ correlated electrons and that the shape
of the peaks depends on the coupling between the quantum dot and the leads.
Furthermore, it is shown that negative differential conductances can occur
due to a 'spin blockade' which is a consequence of the existence of excited
states with different spins. Recently, negative differential conductances
have also been found in the transport through a two--dimensional dot with
parabolic confinement in the fractional quantum Hall effect (FQHE) regime
without spin \cite{kinaret}, where the origin of the effect are excited
states for which the coupling to the leads is weaker than for the ground
state.

As a model, we consider the double barrier Hamiltonian
\begin{equation}\label{h}
H=H_{\rm L}+H_{\rm R}+H_{\rm D}+H_{\rm L}^{\rm T}
+H_{\rm R}^{\rm T}+H^{\rm in} \, ,
\end{equation}
where
$H_{\rm L/R}=\sum_{k}\varepsilon^{\rm L/R}_{k}c^{+}_{{\rm L/R},k}
c^{\phantom{+}}_{{\rm L/R},k}$ describes free electrons in the left/right
lead and
\FL
\[
H_{\rm D}=\sum_{l}(\varepsilon_{l}-e V_{\rm G})c^{+}_{l}c^{\phantom{+}}_{l}
+\! \sum_{l_1,l_2,l_3,l_4}\! V_{l_1l_2l_3l_4}\,
c^{+}_{l_1}c^{+}_{l_2}c^{\phantom{+}}_{l_3}c^{\phantom{+}}_{l_4}
\]
the interacting electrons within the dot. The energies of the
noninteracting electrons are $\varepsilon_{l}$ and
$V_{l_1l_2l_3l_4}$ the matrix--elements of the Coulomb--interaction.
The voltage $V_{\rm G}$ applied to a gate serves to
change the electron density in the well. The barriers are
represented by the tunneling Hamiltonians
$H_{\rm L/R}^{\rm T}=\sum_{k,l}(T^{\rm L/R}_{k,l}c^{+}_{{\rm
L/R},k}c^{\phantom{+}}_{l}+h.c.)$, where $T^{\rm L/R}_{k,l}$ are
the transmission probability amplitudes which we assume to be
independent of $l$. The inelastic term $H^{\rm in}$ allows for
transitions between the dot levels without changing the electron
number. A phononic heat bath and a Fr\"ohlich type coupling
(coupling constant $\sqrt{g}$) would be a microscopic model
leading to such a term. We assume that the phase coherence
between the eigenstates of $H-H^{\rm in}$ is destroyed on a time
scale $\tau_{\Phi}$, which is much larger than the time an
electron needs to travel from one barrier to the other. Thus, the
motion of the electrons inside the dot is sufficiently coherent
to guarantee the existence of quasi--discrete levels. We assume
also that the leads are in thermal equilibrium described by the

Mail/VE V2.0     92196                  read_letter
         1993-06-22         18:13:17     PAGE 3

Fermi--Dirac--distributions $f_{\rm L/R}(\varepsilon )=
(\exp[\beta (\varepsilon -\mu_{\rm L/R})]+1)^{-1}$. The chemical
potential in the left/right lead is $\mu_{\rm L/R}$ and $\beta
=1/k_{\rm B}T$ the inverse temperature. We assume the tunneling
rates through the barriers $t^{\rm L/R}=(2\pi/\hbar)\sum_k|T^{\rm
L/R}_{k,l}|^2 \delta(\varepsilon^{\rm L/R}_{k}-E)$ to be
independent on energy $E$. If they are small compared to the
phase breaking rate $\tau^{-1}_{\Phi}$, the time--evolution of
the occupation probabilities of the many--electron states in the
dot can be calculated using a master--equation.

In contrast to \cite{averin90,averin91}, where changes in the
occupation probabilities for one--electron levels were
considered, we take into account the populations $P_i$ of all
possible Fock states $|i\rangle $ of $H_{\rm D}$. Transitions
between the latter occur when an electron tunnels through a
barrier. Our method allows to determine the stationary non--equilibrium
state without further restrictions. Deviations from
equilibrium linear in the applied voltage have been mentioned in
\cite{beenakker}. In addition the exact many--electron states of
the dot including spin can be taken into account without being
restricted to the conventional charging model. A similar method
was applied in the FQHE regime without spin \cite{kinaret}.

Due to the smallness of $H^{\rm T}$, simultaneous transitions of
two or more electrons \cite{nazarov} which are processes of
higher order in $H^{\rm T}$ are suppressed. Further selection
rules will be specified below. Each of the states $|i\rangle $ is
associated with a certain electron number $n_i$ and with an
energy eigenvalue $E_i$. The transition rates between states
$|i\rangle$ and $|j\rangle $ with $n_{i}=n_{j}+1$ are given by
$\Gamma^{\rm L/R,-}_{j,i}$ and $\Gamma^{\rm L/R,+}_{i,j}$,
depending on whether an electron is leaving or entering the dot
through the left/right barrier, respectively. Here, $\Gamma^{\rm
L/R,-}_{j,i}=t^{\rm L/R}[1-f_{\rm L/R}(E)]$, $\Gamma^{\rm
L/R,+}_{i,j}=t^{\rm L/R}f_{\rm L/R}(E)$ and the electron provides
the energy difference $E=E_{i}-E_{j}$.

Assuming a bosonic heat bath being weakly coupled to the
electrons, the transition rate between $|i\rangle $ and
$|j\rangle $ ($n_i=n_j$) induced by $H^{\rm in}$ is given by
$\Gamma^{\rm in}_{j,i}=r[n_{\rm B}(|E|)+\Theta(E)]$, where $r=g\,
\rho_{\rm ph}$. This is the lowest order result quadratic in the
electron--heat bath coupling strength $\sqrt{g}$. $\rho_{\rm ph}$
is the boson density of states, $n_{\rm B}(E)=(\exp[\beta E]-
1)^{-1}$ the Bose--Einstein--distribution and $\Theta(x)$ the
step function. The matrix of transition rates is
$\Gamma=\Gamma^{\rm L,+}+\Gamma^{\rm R,+}
+\Gamma^{\rm L,-}+\Gamma^{\rm R,-}+\Gamma^{\rm in}$.
The master equation for the
time evolution of the occupation probabilities $P_i$ is
\begin{equation}\label{master}
\frac{d}{dt}P_{i}=\sum\limits_{j\, (j\neq i)}
(\Gamma_{i,j}P_{j}-\Gamma_{j,i}P_{i})\quad,\quad\sum\limits_{i}P_{i}=1\;.
\end{equation}
{}From the stationary solution ($d\bar{P}_{i}/dt=0$) of (\ref{master})
one determines the dc--current
\[

Mail/VE V2.0     92196                  read_letter
         1993-06-22         18:13:17     PAGE 4

I\equiv I^{\rm L/R}=(-/+) e\sum_{i,j\, (j\neq i)}\bar{P}_{j}
(\Gamma^{\rm L/R,-}_{i,j}-\Gamma^{\rm L/R,+}_{i,j})\, .
\]
It equals the number of electrons that pass the left/right barrier per unit
of time.

As a tutorial example we consider the phenomenological charging model
\cite{averin90,averin91,beenakker} for $N$ single--electron levels,
where $\:V_{l_1l_2l_3l_4}=(U/2)\delta_{l_1,l_2}\delta_{l_3,l_4}(1-
\delta_{l_1,l_3})\:$.
The stationary solution of (\ref{master}) was obtained by solving
numerically the system of $2^{N}$ linear equations.

At zero bias voltage the occupation probabilities of the $n$--electron
states are given by a Gibbs distribution
$\:P_{i}^{\rm G}=(\exp [-\beta (E_{i}-\mu n_{i})])/{\cal Z}\:$
with the chemical potential $\mu=\mu_{\rm L}=\mu_{\rm R}$. They solve the
rate equation (\ref{master}) for all $r$. $\cal Z$ is the grand canonical
partition function. For temperatures lower and voltages higher
than the level--spacings, $\bar{P}_{i}$ deviate from equilibrium. For
$r\gg t^{\rm L/R}$ (fast equilibration via bosons) $\bar{P}_i/\bar{P}_j$ can
be satisfactorily approximated by $P_i^{\rm G}/P_j^{\rm G}$ for $n_i=n_j$.
This can be seen in Fig.~\ref{pe} where $\ln \bar{P}_{i}$ for a given
$n_{i}$ lie on straight lines with slope $-\beta $. This confirms the
assumptions of a Gibbs distribution among states with given electron
number in \cite{averin91,beenakker}.
When $n_i\ne n_j$, $\bar{P}_i/\bar{P}_j$ can be far from equilibrium.
It is impossible to scale all of the points onto one common curve
by defining an effective chemical potential for the dot \cite{nonlin}.

The current--voltage characteristics (Fig.~\ref{iv}) for temperatures lower
than the level--spacing shows finestructure in the Coulomb staircase
consistent with recent experiments \cite{johnson} and earlier theoretical
predictions using a different approach \cite{averin90}. Intra--dot
relaxation ($\sim r$) suppresses the lowest of the finestructure steps
because the electron that contributes to the current at the $n$--th Coulomb
step has to enter the $n$--th or a higher one--electron level. For $r\gg
t^{\rm L/R}$ the $n-1$ other electrons occupy with high probability all of
the lower one--electron levels. Asymmetric coupling to the leads changes the
height of the steps in the $I$--$V$ curve. This can be explained for the
$n$--th Coulomb step as follows. If $t^{\rm L}>t^{\rm R}$
($\mu_{\rm L}>\mu_{\rm R}$) the stationary occupation probabilities
favor the $n$--electron levels, while for $t^{\rm L}<t^{\rm R}$
the $(n-1)$--electron states are preferred. Since there are more
$n$--electron levels than $(n-1)$--electron levels, the probability for an
electron to escape is reduced in the former case as compared to the
probability for an electron to enter in the latter case. These processes
limit the current. They lead to a reduction and an enhancement of the
current in the first and second case, respectively.

For fixed $V$, the conductance shows peaks when $V_{\rm G}$ is
varied. The linear response limit of our method is in agreement
with \cite{meir}. For finite bias voltage,
$eV=\mu_{\rm L}-\mu_{\rm R}$, larger than the level spacing, transitions
involving excited states can occur. The number of levels that
contribute to the current varies when $V_{\rm G}$ is changed.
This leads to the splitting of the conductance peaks observed
experimentally and explained qualitatively in

Mail/VE V2.0     92196                  read_letter
         1993-06-22         18:13:17     PAGE 5

\cite{johnson,foxman}. From the quantitative treatment of the
charging model using (\ref{master}) for $T=0$, i.e. only constant
nonvanishing or vanishing $\Gamma_{i,j}$'s, we obtain that the
number of transitions contributing to the current varies with
$V_{\rm G}$ as $0-6-4-12-4-6-0$ in the specific example shown for
finite temperature in Fig.~\ref{iv}, inset. Taking into account
the stationary $\bar{P}_i$'s the sequence of current values is
$0-3/2-4/3-2-4/3-3/2-0$. If the difference $E_{0}(n)-E_0(n-1)$
between the energies of the many electron ground states lies
outside the interval $[\mu_{\rm R},\mu_{\rm L}]$ the transport
via other energetically allowed transitions is Coulombically
blocked. While the relaxation rates have almost no influence on
the conductance, asymmetric coupling to the leads changes the
shape of the peaks considerably. We propose to explain the slight
asymmetry observed in the experiment \cite{johnson} by the
asymmetry of the barriers and we predict that the asymmetry in
the finestructure of the observed conductance peaks will be
reversed if the sign of the bias voltage is changed. Such
asymmetric conductance properties can be used to construct a
mesoscopic rectifier. Similar effects were inferred earlier from
the high frequency properties of mesoscopic systems containing
asymmetric disorder \cite{falko}.

However, the charging model is a severe simplification for
interacting electrons, especially for systems with reduced
dimensionality. Therefore, we consider as a second example $n\le 4$
interacting electrons in a quasi one--dimensional (1D) square
well of length $L$ including the spin degree of freedom
\cite{corr92}. We calculated numerically the exact eigenvalues
$E_{\nu}$ and the corresponding $n$--electron states $|\nu\rangle$
for this correlated electron model. The interaction potential
$\propto ((x-x')^2+\lambda^2)^{-1/2}$ was used, where $\lambda$
($\ll L$) is due to a transversal spread of the electronic wave
function. Since the interaction is spin independent, the $n$--electron
spin $S$ is a good quantum number. The properties of the
correlated states and the energy spectrum are discussed in detail
in \cite{corr92}. For not too large electron densities tendency
towards Wigner crystallization is found. In this regime, the
excitation spectrum consists of well separated multiplets, each
containing 2$^n$ states. The energetic differences between
adjacent multiplets decrease algebraically with electron density.
They correspond to vibrational excitations. The considerably
smaller intra--multiplet energy differences decrease
exponentially. The wave functions of individual levels within a
given multiplet differ in symmetry and $S$. The excitation
energies in the lowest multiplet can be calculated analytically
\cite{pocket} and depend only on one tunneling integral $t_n$
(Table \ref{table}). In summary, two different energy scales
characterize the $n$--electron excitations. We will now
demonstrate that they can be distinguished in principle by a
nonlinear transport experiment.

As an additional selection rule, we take into account that each
added or removed electron can change the total spin $S$ of the
$n$ electrons in the dot only by $\pm 1/2$ with probabilities
$(S+1)/(2S+1)$ and $S/(2S+1)$, respectively. We emphasize here
again that this can be done only by considering all possible Fock
states in the rate equation (\ref{master}).

Mail/VE V2.0     92196                  read_letter
         1993-06-22         18:13:17     PAGE 6

Occupation probabilities are similar as for the charging model
but modified by spin effects. Current--voltage characteristics
and conductivity peaks calculated by using the excitation
energies given in Table \ref{table} are shown in Fig~\ref{ivex}.
First of all, we observe that the lengths of the steps in the
Coulomb staircase and accordingly the distances of the conductivity
peaks are no longer equal since the exact $n$--electron ground
state energy is not proportional to $n(n-1)$ as
in the charging model for small $\varepsilon_{l}$'s. The
deviation from the classical behavior is related to the
inhomogeneity of the quantum mechanical charge density of the
ground state \cite{corr92}. Second, the heights of the
finestructure steps are more random as compared to those in
Fig.~\ref{iv} due to the non--regular sequence of total spins
(cf.\ Table \ref{table}) and the spin selection rules. In certain
cases finestructure steps in the $I$--$V$ characteristic may even
be completely suppressed.

Strikingly, regions of negative differential conductance occur
(Fig.~\ref{ivex}). They are related to the reduced possibility
for the states of maximal spin $S=n/2$ to decay into states of
lower electron number. The states with maximal spin occur only
once within each multiplet for given electron number. Therefore
only one finestructure step with negative differential
conductance can occur within each Coulomb step. The peak in the
$I$--$V$ curve can become less pronounced if $t^{\rm L}<t^{\rm R}$,
because then the dot is empty and the $(n-1)\rightarrow n$
transitions determine the current. The spin selection rules
reduce the probability for $n\rightarrow (n-1)$ transitions
(especially important for $t^{\rm L}>t^{\rm R}$) and the negative
differential conductance becomes more pronounced
(Fig.~\ref{ivex}). Such a behavior can in fact be seen in the
experimental data \cite{johnson} but certainly needs much more
elaborate further investigations. These negative differential
conductances can in principle be used to construct a mesoscopic
oscillator.

In summary, we have investigated nonlinear transport through a
double barrier taking into account Coulomb interactions, spin and
non--equilibrium effects. For two model Hamiltonians occupation
probabilities, current--voltage characteristics and conductances
versus gate--voltage at finite bias voltage have been calculated
using a master equation approach.

Thermally induced intra--dot relaxation processes lead to a
suppression of the $n$ lowest finestructure steps in the $n$--{th}
Coulomb step of the $I$--$V$ curve. At finite bias voltages,
the intra--dot relaxation results in thermal equilibrium only
among the states with equal dot electron number. We have
demonstrated explicitly that the stationary non--equilibrium
populations cannot be described by a Gibbs distribution.
Asymmetric barriers cause pronounced asymmetries in the
conductance peaks versus gate--voltage. We predict the reversal
of the asymmetry when the bias voltage is reversed. Taking into
account the quantum mechanics of Coulombically interacting
electrons including their spins leads to striking modifications
of the transport as compared to the charging model. First of all,

Mail/VE V2.0     92196                  read_letter
         1993-06-22         18:13:17     PAGE 7

the Coulomb blockade intervals in the $I$--$V$ characteristic and
the distances between the conductance peaks are no longer
constant. Because of spin selection rules for the correlated
electron system, the heights of the single steps in the
finestructure of the Coulomb staircase look random. Furthermore,
regions of negative differential conductance occur because for
each electron number the one state of maximum spin has a reduced
transition probability into states with lower electron number.
This general feature of a 'spin blockade' is not restricted to
the quasi--1D model considered here but should also apply to 2D
dots used in experiments containing few electrons. All of the
theoretically predicted features described above are
qualitatively consistent with experiment \cite{johnson}. Further
experiments, in particular using `slim quantum dots', are however
necessary in order to clarify the quantitative aspects.

Preliminary results taking into account a magnetic field in
transport direction show that the occurrence of negative
differential conductance is influenced but not always suppressed.
To clarify these questions and to be able to make quantitative
comparisons with existing experimental data, generalization of
the above correlated electron model to 2D is necessary.

For valuable discussions we thank R.~Haug and J.~Weis.
This work was supported in part by the Deutsche Forschungsgemeinschaft via
grants We 1124/2--2, We 1124/4--1 and AP 47/1--1 and by
the European Community within the SCIENCE program,
grant SCC$^{*}$--CT90--0020.

\begin{figure}
\caption[fig1]{Stationary occupation probabilities $\bar{P}_{i}$
for a dot containing $N=6$ one--electron levels. Electron numbers
are $n_{i}=1$ ($\Box $), $n_{i}=2$ ($\triangle $), $n_{i}=3$
($\Diamond $) and $n_{i}=4$ ($\bigcirc $). $t^{\rm L}=t^{\rm R}$,
$\mu_{\rm L}=1.5U$, $\mu_{\rm R}=-0.3U$ and $\mu=(\mu_{\rm L}
+\mu_{\rm R})/2$. Energies of the one--electron levels are
$\varepsilon_{1}=0.1U$, $\varepsilon_{2}=0.2U$,
$\varepsilon_{3}=0.3U$, $\varepsilon_{4}=0.4U$,
$\varepsilon_{5}=0.5U$ and $\varepsilon_{6}=0.6U$. Inverse
temperature is $\beta =25/U$ and the relaxation rate $r=100\bar{t}$.
$\bar{t}=t^{\rm L}t^{\rm R}/(t^{\rm L}+t^{\rm R})$ is the
total transmission rate.\label{pe}}
\end{figure}

\begin{figure}
\caption[fig2]{Current--voltage characteristic of a dot
represented by $N=6$ one--electron levels. Model parameters are
as in Fig.~\ref{pe}. Inverse temperature is $\beta=100/U$,
$\mu_{\rm R}=0$. Inset: Current versus $V_{\rm G}$ for $\mu_{\rm L}
=0.26 U$ and $\mu_{\rm R}=0$. $V=0.26U/e$ is between the double
and the triple of the bare level--spacing such that the
conductance peaks are modulated as explained in the text. Dashed
lines: results for $r=0$ and equal barriers, dotted and solid
lines: $t^{\rm R}/t^{\rm L}=0.5$ and $2$, respectively. Shaded
regions: suppressions of steps induced by relaxation
$r/\bar{t}=100$ at $t^{\rm L}=t^{\rm R}$.\label{iv}}
\end{figure}

\begin{figure}
\caption[fig3]{Current--voltage characteristic ($\mu_{\rm R}=0$)
and the splitting of the fourth conductance peak at $\mu_{\rm
L}=0.3 E_{\rm H}$ and $\mu_{\rm R}=0$ (inset) of a dot described
by the correlated electron model for $\beta=200/E_{\rm H}$
($E_{\rm H}\equiv e^{2}/a_{\rm B}$ Hartree--energy) and
$r=\bar{t}$. Tunneling integrals are $t_{2}=0.03E_{\rm H}$,
$t_{3}=0.07E_{\rm H}$ and $t_{4}=0.09E_{\rm H}$, numerically
determined ground state energies $E_{0}(1)=0.023E_{\rm H}$,
$E_{0}(2)=0.30E_{\rm H}$, $E_{0}(3)=0.97E_{\rm H}$,
$E_{0}(4)=2.15E_{\rm H}$. Dashed, dotted and solid lines
correspond to $t^{\rm R}/t^{\rm L}=1$, $0.5$ and $2$,
respectively.\label{ivex}}
\end{figure}

\begin{table}
\caption{Spin and energies of low lying excitations of the
correlated electron model at sufficiently large electron
distances $r_{\rm s}\equiv L/(n-1)\gg a_{\rm B}$. The tunneling
integrals $t_n$ decrease exponentially with $r_{\rm s}$.}
\begin{tabular}{ccc}
$n$&$S$&$E_{\nu}-E_{0}(n)$\\ \tableline
$2$&$0$&$0$\\
$2$&$1$&$2t_2$\\
$3$&$1/2$&$0$\\
$3$&$1/2$&$2t_3$\\

Mail/VE V2.0     92196                  read_letter
         1993-06-22         18:13:17     PAGE 9

$3$&$3/2$&$3t_3$\\
$4$&$0$&$0$\\
$4$&$1$&$(1-\sqrt{2}+\sqrt{3})t_4$\\
$4$&$1$&$(1+\sqrt{3})t_4$\\
$4$&$0$&$(2\sqrt{3})t_4$\\
$4$&$1$&$(1+\sqrt{2}+\sqrt{3})t_4$\\
$4$&$2$&$(3+\sqrt{3})t_4$\\
\end{tabular}
\label{table}
\end{table}


\begin{references}
\bibitem[b]{Addr} Address to which correspondence should be directed.
\end{references}
\end{document}